\documentclass[twocolumn,amsmath,prl,showpacs]{revtex4}

\input{epsf}
\epsfclipon

\begin{document}

\title{Directional Ordering of Fluctuations in a Two-dimensional Compass Model}

\author{Anup Mishra$^1$, Michael Ma$^1$, Fu-Chun Zhang$^{1,2}$,
Siegfried Guertler$^{3,4}$, Lei-Han Tang$^3$, and Shaolong Wan$^{3,5}$}
\affiliation{$^1$Department of Physics, University of Cincinnati,
Cincinnati, Ohio 45221-0011\\
$^2$Department of Physics, University of Hong Kong, Pokfulam Road,
Hong Kong SAR, China\\
$^3$Department of Physics, Hong Kong Baptist University, Kowloon Tong,
Hong Kong SAR, China\\
$^4$Institute for Theoretical Physics, Technical University Graz,
Petersgasse 16, A-8010 Graz, Austria\\
$^5$Department of Modern Physics, University of Science and Technology of
China, Hefei, China
}

\date{\today}

\begin{abstract}
In the Mott insulating phase of the transition metal oxides, the effective
orbital-orbital interaction is directional both in the orbital space and
in the real space. We discuss a classical realization of directional
coupling in two dimensions. Despite extensive degeneracy of the ground
state, the model exhibits partial orbital ordering in the form of directional ordering of fluctuations at low temperatures
stabilized by an entropy gap. Transition to the disordered phase is shown
to be in the Ising universality class through exact mapping and
multicanonical Monte Carlo simulations.

\end{abstract}


\maketitle

Recently, there has been growing interest in the effects of orbital
degeneracy in the physics of transition metal oxide
insulators\cite{{khomskii},{nagaosa}}. In these systems,
the dominating energy scales for d-electrons on transition
metal (TM) ions are the on-site Coulomb repulsion (which freezes out the
charge degrees of freedom), the Hund's rule coupling, and the crystal field
due to the surrounding oxygen ions. The latter two together determine the
degeneracies and degrees of freedom of spin and orbital on each
transition metal ions.
Spins and orbitals on neighboring TM ions can then be coupled through the
superexchange mechanism. In the case of orbitals, they are also coupled
through the phonon-mediated cooperative Jahn-Teller mechanism\cite{khomskii}. These couplings
determine the low temperature properties of these systems.

Because orbital coupling is intrinsically directional\cite{mostovoy}, orbital ordering
brings up some unusual questions. Especially interesting is when the coupling along a given bond
direction is Ising like, but with different Ising axes along different bond
directions \cite{khomskii2}. The Hamiltonian for orbitals is then given by
\begin{equation}
H=-J\sum_{\left\langle ij\right\rangle }\mathbf{\tau }_{i}\cdot \widehat{%
\mathbf{n}}_{ij}\mathbf{\tau }_{j}\cdot \widehat{\mathbf{n}}_{ij}
\label{d_orb_H}
\end{equation}
where $\mathbf{\tau}$ is an isospin operator representing the orbital degree
of freedom, and $\widehat{\mathbf{n}}_{ij}$ is a unit vector giving the
Ising axis for the bond $\left\langle ij\right\rangle .$ For example, for $%
e_{g}$ orbitals in Perovskite structures, $\widehat{\mathbf{n}}_{ij}$ for the
three different bond directions are coplanar and oriented relative to each
other by $120^\circ,$ giving rise to the so-called $120^\circ$
Model\cite{{khomskii2},{nussinov},{biskup}}.
This model
is also applicable to $t_{2g}$ orbitals on the three bonds of a honeycomb
lattice as in planes of $V_{2}O_{3}$\cite{{castellani},{joshi}}.
On the other hand, for $t_{2g}$
orbitals on Perovskite structures, the relevant model is the Compass Model\cite{{khomskii}, {khomskii2},{maekawa}}
with $\widehat{\mathbf{n}}_{ij}=\widehat{x},$ $\widehat{y},$ $\widehat{z}$
for the three bond directions. The Compass Model may also feature as part of the
spin-spin coupling of $t_{2g}$ orbitals when spin-orbit interaction is taken
into account\cite{khaliullin}. A common feature of both the Compass Model and the $120^\circ$ Model is
the competition between bonds in different directions, with the resulting frustration leading to macroscopic degeneracy of the classical ground state.

In this Letter, we report analytical and numerical results on a
classical version of (\ref{d_orb_H}) in 2D. The highly anisotropic coupling gives rise to interesting interplay between
one- and two-dimensional ordering, between continuous and discrete spin physics, and between slow and fast modes. Our main result is that at
low $T>0,$ there is no conventional ordering\cite{lawson}, but there is nevertheless LRO in the form of a directional ordering in
fluctuations. This ordering corresponds to a partial breaking of the 4-fold symmetry and is stabilized by entropy.
In this phase, the system exhibits spontaneous dimension reduction by essentially decoupling into one-dimensional (1D)
chains running either horizontally or vertically along the bonds.
Through exact mapping and extensive Monte Carlo simulations, we show 
that this ordering transition belongs
to the Ising universality class. We then discuss generalizations of
these results to the quantum case and in three dimensions, as well as their implications with respect to orbital ordering.

Consider the classical Compass model on a square lattice
of $N=L\times L$ sites,
\begin{equation}
H=-J\sum_{i}(S_{ix}S_{i+\widehat{x},x}+S_{iy}S_{i+\widehat{y},y}),
\label{compass-H}
\end{equation}
where $\mathbf{S}_{i}=(\cos \theta _{i,}$ $\sin \theta _{i})$
represents either a real spin or an orbital isospin. On the square lattice, the sign of $J$ can be gauged away, so we take $J>0.$ Along each row (column), we have a simple Ising model (IM) with
quantization axis along $\widehat{x}(\widehat{y}).$
Hamiltonian (\ref{compass-H}) has two types of discrete symmetry.
I) There is a global four-fold
rotation symmetry corresponding to simultaneously rotating the spins and
lattice by multiples of $90^\circ.$ II) In addition, it is also invariant
under the 1D spin flip transformation $S_{ix}\rightarrow -S_{ix},$
$S_{iy}\rightarrow S_{iy}$ for all $i$ on any one row and
$S_{ix}\rightarrow S_{ix},$
$S_{iy}\rightarrow -S_{iy}$ for all $i$ on any one column.
Since symmetry I is two-dimensional (2D), we expect that it may
be broken at finite $T,$ while the 1D nature of symmetry II should
imply no symmetry breaking except possibly at $T=0.$ We will see indeed that
this is the case, but the physics leading to it and their consequences are
not trivially deduced from such symmetry considerations.

The low temperature properties of (\ref{compass-H}) are further complicated
by an additional $O(2)$ degeneracy of the ground state.
Apart from a constant term, Eq. (\ref{compass-H}) can be written as,
\begin{equation}
H=\frac{J}{2}\sum_{i}\left[(\cos \theta_{i}-\cos\theta_{i+\widehat{x}})^{2}
+(\sin\theta_{i}-\sin\theta_{i+\widehat{y}})^{2}\right].
\label{H-sym}
\end{equation}
Clearly $\theta_{i}\equiv 0$ is a ground state, as are the
$D=2\times 2^{L}$ states obtained from it by the symmetry operations I and II. However, Eq.~(\ref{H-sym})
shows that the ground state energy is invariant under arbitrary global
rotation of $\theta _{i}\equiv 0$ to $\theta _{i}\equiv \theta$. Unlike the
isotropic XY model, where this invariance holds for each bond, here the
energy loss from the horizontal bonds are compensated by energy gain in the
vertical bonds. Thus, the ground state exhibits an $O(2)$ degeneracy not
related to the symmetries of $H.$ We will see that this ``accidental''
degeneracy is lifted at finite temperatures by entropy due to slow mode fluctuations.

Upon a redefinition of the spins through symmetry II, any of the ground states mentioned above can be recast as $\theta _{i}\equiv \theta.$ To study slow mode physics, we start with the spin wave or harmonic approximation. 
Expanding (\ref{H-sym}) to the second order in
$\varphi_i= \theta_i-\theta$, we obtain the spin wave Hamiltonian
in Fourier form,
$H_{\rm SW}=J\sum_{k}\epsilon_k(\theta)\left| \varphi_{k}\right|^{2}$.
For general $\theta,$ the spin wave spectrum
$\epsilon_k(\theta)=(1-\cos k_{x})\sin^2\theta
+(1-\cos k_{y})\cos^2\theta$ is an anisotropic 2D one, with
zero modes at $k_x=k_y=0$. However,
for special ordering directions $\theta =0,$ $\pi /2,$ $\pi ,$ and $3\pi /2,$
the spectrum becomes 1D like, independent of either $k_{x}$ ($\theta =0,$ $%
\pi $) or $k_{y}$ ($\theta =\pi /2,$ $3\pi /2$), and the density of states
of gapless excitations is 1D rather than 2D. The high density of
low lying states suggests an entropic mechanism to stabilize these four
directions at $T>0$.

To put the discussion on a firmer footing, we employed a
self-consistent harmonic approximation for the ordered phase at $\theta =0.$
Based on the Bogolyubov-Peierls theorem,
$F\leq F_{0}-\langle H_{0}\rangle_{0}+\langle H\rangle_{0},$
we compute the variational free energy using a trial Hamiltonian
$H_{0}=J\sum a_{k}\left| \varphi_{k}\right|^2$. Minimizing
the free energy, we obtain
\begin{equation}
a_{k}=m+\gamma _{x}(1-\cos k_{x})+\gamma _{y}(1-\cos k_{y})
\end{equation}
where $m$ is the self-consistent spin-wave gap, and $\gamma _{x}$ and $%
\gamma _{y}$ are the self-consistent stiffness.

At low temperatures,
a 1D spectrum with $\gamma_x(T)=0,$ $\gamma_y(T)=1-O(T^{2/3}),$
and $m(T)=\frac{1}{2}T^{2/3}+O(T)$ is obtained.
Within the
self-consistent approach, anharmonic effects are incorporated into a shift
of these parameters at finite $T$ from their bare spin wave values.
Most significantly we see that a gap $m$ is generated, which suppresses the
diverging 1D fluctuations in the spin wave analysis, and stabilizes the ordering along one of the 4 special directions.

To address whether there is ordering into one of the D degenerate ground states, we need to consider the effects of fast modes, or more precisely, abrupt spin
flips. For this purpose, the continuous nature of the spins should not be
crucial, so we discretize the Compass model into a ``4-state Potts Compass
model'' with the same symmetry as (\ref {compass-H}), given by
\begin{equation}
H_{\rm P}=-J\sum (n_{i\sigma }n_{i+\widehat{x},\sigma }
\sigma_{i}\sigma_{i+\widehat{x}}
+n_{i\tau}n_{i+\widehat{y},\tau }\tau_{i}\tau _{i+\widehat{y}}),
\label{Potts}
\end{equation}
where on each site we have ``occupation numbers'' $n_{i\sigma }=0,1$ and $%
n_{i\tau }=1-n_{i\sigma }.$ If $n_{i\sigma }=1,$ then there is an additional
internal degree of freedom $\sigma =\pm 1;$ and similarly for $n_{i\tau}$
and $\tau$. The correlation of these internal degrees of freedom with the
occupation numbers together with the constraint in the latter couple these
various variables.

The partition function of the Potts compass mode takes the form
$Z_{\rm P}={\rm Tr}_{\left\{ n_{i\sigma }\right\} }
{\rm Tr}_{\left\{ \sigma _{i},\tau_{i}\right\} }^{\prime }
\exp (-\beta H_{\rm P})$,
where ${\rm Tr}^{\prime }$ indicates that for a given configuration of
$\left\{n_{i\sigma }\right\}$ the trace over $\sigma (\tau )$ should
only be on those sites with $n_{i\sigma}=1(0)$.
On the other hand, we note that if for
example, $n_{i\sigma }=0,$ then $H_{\rm P}$ is independent of $%
\sigma _{i}$, and tracing over $\sigma _{i}=\pm 1$ simply gives a superfluous
factor of $2.$ Thus, ${\rm Tr}^{\prime }$ can be replaced by the
unrestricted ${\rm Tr}$ to give
$Z_{\rm P}={2^{-N}}{\rm Tr}_{\left\{ n_{i\sigma }\right\} }
{\rm Tr}_{\left\{
\sigma _{i},\tau _{i}\right\} }\exp (-\beta H_{\rm P})$.
The trace over $\sigma $
and $\tau $ can now be easily done using transfer matrix since $H_{\rm P}$
consists of decoupled 1D chains as far as $\sigma $ and $\tau $ are concerned,
resulting in
$Z_{\rm P}={\rm Tr}_{\left\{ n_{i\sigma }\right\} }\exp (-\beta H_{\rm eff}),$
where
\begin{eqnarray}
H_{\rm eff}&=&-T\ln[\cosh(\beta J)]\sum_{i}
(n_{i\sigma }n_{i+\widehat{x},\sigma}
+n_{i\tau }n_{i+\widehat{y},\tau })\nonumber\\
&&
-T\ln[1+\tanh ^{L}(\beta J)]
\bigl[\sum_{\alpha }C_{\alpha }+\sum_{\gamma }D_{\gamma}\bigr],
\label{H_eff}
\end{eqnarray}
In the last two terms of $H_{\rm eff}$, $C_{\alpha }=\prod_{i}n_{i\sigma }n_{i+\widehat{x},\sigma},$ for 
all sites $i$ in the row
$\alpha ,$
while $D_{\gamma }=\prod_{j}n_{j\tau
}n_{j+\widehat{y},\tau },$
for all sites $j$ in the column $\gamma .$ Under periodic boundary conditions, $C_\alpha=1$ if $n_{i\sigma}=1$
for all sites in row $\alpha$ and zero otherwise. Similarly,
$D_\gamma=1$ if $n_{i\tau}=1$ for all sites in column $\gamma$ and
zero otherwise. At $T=0$, we recover the $2\times 2^{L}$-fold
degeneracy of the ground state due to symmetries I and II.

At any $T>0$, the last two terms in Eq.~(\ref{H_eff}) are finite-sized terms that vanish in the
thermodynamic limit $L\rightarrow\infty$.
Ignoring them, we may rewrite $H_{\rm eff}$ in terms of $n_{i\sigma ,\tau }=\frac{1}{2}(1\mp \mu _{i})$ as
\begin{equation}
H_{\rm eff}=-2N\tilde J-\tilde J\sum_{i}(\mu_{i}\mu_{i+\widehat{x}}
+\mu_{i}\mu_{i+\widehat{y}}).
\label{Potts-Ising}
\end{equation}
The 4-state Potts Compass model is thus mapped exactly into the 2D Ising Model (2DIM).
The coupling constants of the two models are related by
$\tilde J=T\ln[\cosh (J/T)]/4.$
From the 2DIM exact $\tilde T_{c}=2\tilde J/\ln(1+\sqrt{2})$,
we conclude that the Potts Compass
model has long-ranged order (LRO) for all $T<T_c=0.4084J$. What is the nature of this LRO? First, note that because the
trace over $\sigma $ and $\tau $ are for decoupled chains, $\left\langle
\sigma _{i}\right\rangle $ and $\left\langle \tau _{i}\right\rangle \equiv 0$
for all $T>0.$ Instead, the 2DIM uniform ordering of $\left\langle \mu
_{i}\right\rangle $ corresponds to $\left\langle n_{\sigma }\right\rangle
-\left\langle n_{\tau }\right\rangle \neq 0.$ In other words, the ordering
is not of the spins spontaneously pointing along one of the 4 possible
states, but of them having stronger fluctuations in 2 of the 4 states,
henceforth called directional ordering of fluctuations.
In this phase, the $Z_{4}$ symmetry of the Compass model is only
partially broken into $Z_{2}\times Z_{2}.$ While at $T=0$, the ground state has macroscopic degeneracy, the free energy has only two degenerate minima at $T=0_+$

Based on symmetry considerations and on our entropy stabilization arguments earlier, we expect the above conclusions to hold also for the continuous Compass model except for the value of $T_{c}$. To confirm this and to rule out a preemptive first order transition, we perform Monte Carlo simulations. However, such simulations are complicated by the strong size dependence that originates
from the finite-sized terms in Eq. (\ref{H_eff}) under periodic boundary
condition when the 1D correlation length $\xi_{\rm 1D}$ exceeds the linear system size $L$ at sufficiently low temperature. This boundary contribution, however, can be eliminated
if, for each row and column, we pick one bond and allow the sign of
the coupling constant on that bond to be an additional degree of freedom.
The data presented below are
obtained under this ``annealed'' boundary condition. In view of the
large number of degenerate low energy states arising from Symmetry II,
the simulation is performed using the multicanonical Monte Carlo sampling
scheme\cite{Berg92}. Details will be published elsewhere.

Figure 1(a) shows the average of the directional bond order parameter
$q=N^{-1}\sum_i(S_{ix}^{2}-S_{iy}^{2})=N^{-1}\sum_i \cos 2\theta_{i}$
against $T$ for $L=8,12,16,24,32$ and 48.
Existence of an ordered state at low temperatures is evident from the data.
To locate the transition point, we computed the Binder cumulant
$B=1-\langle q^4\rangle/(3\langle q^2\rangle^2)$ for various
system sizes, as shown in Fig. 1(b). From the crossing of the
curves, we estimate $T_c=0.147\pm 0.001$ for the continuous
Compass model. The value of $B$ at the crossing point for
large system sizes also agrees reasonably well with the 2DIM
result $B_c=0.61069\ldots$.\cite{Binder_c}

\begin{figure}
\epsfxsize=8cm
\epsfbox{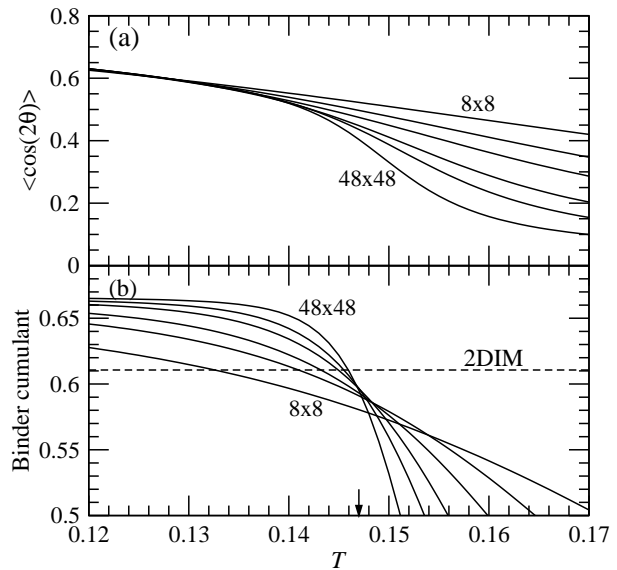}
\caption{(a) Directional bond ordering parameter against temperature
$T$ for the 2D Compass model at various system sizes.
(b) Binder cumulant against temperature.
The critical temperature is estimated at $T_c=0.147\pm 0.001$.
}
\label{order_para}
\end{figure}

Figure 2 shows the specific heat data for the six different sizes up to
$L=48$. A weak divergence near the $T_c$ value determined above is
clearly seen. The inset shows the same data in the critical region after subtraction of a background linear function,
plotted using the scaled variables according to the finite-size
scaling form derived in Ref.\cite{fisher-ferdinand}.
Apart from the smallest size at $L=8$, the data collapse is quite
satisfactory.

\begin{figure}
\epsfxsize=8cm
\epsfbox{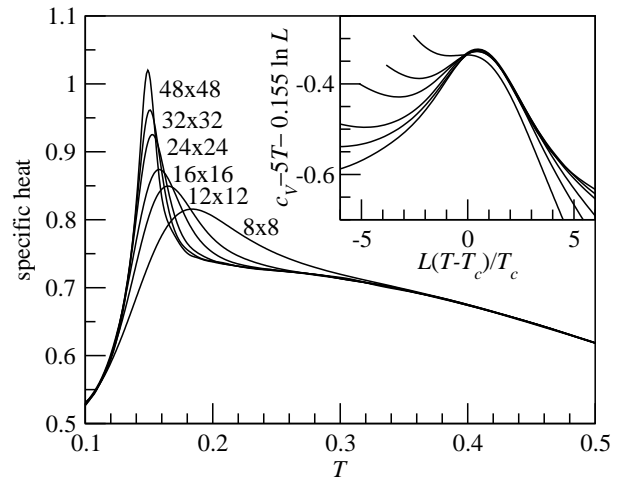}
\caption{Specific heat against temperature for the 2D Compass model
under the annealed boundary condition. Inset: same data plotted
using scaled variables.
}
\label{specific_heat}
\end{figure}

The simulation data shows quite convincingly that directional bond
ordering exists at low temperatures, and the transition to the disordered
phase is in the Ising universality class.
The transition temperature of the continuous Compass model, on the
other hand, is considerably lower than that of the Potts Compass model.
We attribute this to the softening of domain wall energy
in the continuous model. Indeed, the free energy gap between the favored orientation
and other states does not remain constant, but in fact vanishing as
$T^{2/3}$ at low $T$.

We next discuss the implications of our results to orbital ordering, taking
the spin in the Hamiltonian to represent the orbital isospin. For that we
should take into account these spins will then have $S<\infty $, i.e.
quantum mechanical rather than classical, and so there will be fluctuations
about the classical ground state even at $T=0.$ For TMO, the relevant $S$ are $1/2$ and $1$ for double and triple orbital degeneracy respectively. For slow modes, the physics
is somewhat similar to that of the Heisenberg antiferromagnet on the Kagome
lattice, and also similar to straightforward extension of the finite
temperature classical analysis above to $2+1$ dimensions, with renormalized
spin wave fluctuations lifting the $O(2)$ degeneracy while also generating a
gap that stabilizes the $\theta =0,$ $\pi /2,$ $\pi ,$ and $3\pi /2$ ordered
states. The same $2+1$ dimension argument would imply that the fast mode
physics would {\it {not}} destroy the conventional ordering at $T=0.$ However, the
true description of the fast mode physics would involve understanding the
physics of instanton tunneling of the quantum Compass Model, and is beyond
the scope of this paper. At any rate, at low but finite $T,$ we expect both the
classical and quantum mechanical models to exhibit directional but not
conventional ordering, thereby leaving the orbital degeneracy completely ($S=1/2$) or partially ($S=1$) unbroken. In the case of $S=1$, the directional ordering reduces the triple orbital degeneracy to double degeneracy. In the case of $S=1/2,$
$S_{ix}^{2}=S_{iy}^{2}\equiv 1/4,$ and the directional order parameter $q$ defined above $\equiv 0.$ Instead, we define an alternative
directional order parameter applicable for all $S$, namely, $r=<S_{ix}S_{i+x,x}-S_{iy}S_{i+y,y}>,$ which like $q$ shows LRO in the
direction of fluctuations in isospin space. An advantage of $r$ over $q$ is that it explicitly displays the energy difference
between horizontal and vertical bonds and hence the broken lattice rotation symmetry.
A consequence of this is that,
when the couplings of the orbital isospin to
lattice modes are included, the directional ordering will be necessarily
accompanied by a lattice distortion so that the bond length in horizontal
and vertical directions become unequal.

Our results can be generalized to the Compass model on the 3D cubic lattice,
with now a 3-component spin $\mathbf{S}=(S_{x},S_{y},S_{z}).$ The slow mode
physics is basically identical to the 2D case, so that orderings along $\pm
\widehat{x},$ $\pm \widehat{y},$ $\pm \widehat{z}$ have 1D spin wave
spectrum within harmonic approximation. Anharmonic terms then generate an
entropy gap as in the 2D case. Restricting to the 6 special directions, we
can then consider the corresponding 6-state Potts compass model{\cite{3d}}.
The mapping
we used to map the 2D\ 4-state Potts Compass model into the 2D Ising model
can be applied here too. However, in this case, the mapping does not result
in an ordinary 3-state Potts model, but in a 3-state Potts Compass model
$H=-\sum_{i}n_{i\sigma }n_{i+x,\sigma }+n_{i\tau }n_{i+y,\tau }+n_{i\mu
}n_{i+x,\mu }$,
where $n_{i\sigma ,\tau ,\mu }=0,1$ and $n_{i\sigma }+n_{i\tau }+n_{i\mu
}=1. $ There is no known solution to this model. Heuristic domain wall
analysis suggests that directional ordering is stable at low $T,$ but more
rigorous calculation is necessary for this to be conclusive. Unlike the 
$2D$ case, the degeneracy of the directional ordered state, while less than 
the ground state degeneracy of $D=3\times (2\times 2^{L})^{L}$, is also 
macroscopically degenerate. For example, beginning with the directional 
ordered state $n_{i\sigma} \equiv 1,$ one can take all the spins on an 
arbitrary number of planes perpendicular to the z-axis and convert them all to 
$n_{i\tau}=1$ with no cost in free energy. The degeneracy is then found to
be
$3\times 2\times 2^{L}$. 
Notice that the
fluctuation effects are stronger in 3D than 2D, and in fact, in general, the
higher the dimension, the stronger the fluctuations, a point previously
pointed out by Khomskii and Mostovoy in Ref.\cite{khomskii2}.

In conclusion, we have established that the Compass model has a low 
temperature phase characterized by 
$\left\langle \mathbf{S}_{i}\right\rangle = 0.$ but with long-ranged 
correlations in the direction of fluctuations in both isospin and 
lattice space.

We thank J. Perk, H. Au-Yang Perk, H. G. Evertz, and G. Khaliullin 
for useful discussions. SG thanks the Physics Department, Hong Kong Baptist
University (HKBU) for hospitality. Research is supported in part by
the Research Grants Council of the Hong Kong SAR under grants
HKBU 2061/00P and HKBU 2017/03P, and by the HKBU under grant
FRG/01-02/II-65 (SG, LHT, SW).
Computations were carried out at HKBU's High
Performance Cluster Computing Center supported by Dell and Intel.


\begin{thebibliography}{99}
\bibitem{3d}In the case of the quantum model, there is the
additional complication that unlike the case of 2D, symmetry II holds only
classically.
\end{thebibliography}
\end{document}